\documentclass[reprint,prx,aps,amsmath,amssymb,floatfix,superscriptaddress,longbibliography]{revtex4-2}
\usepackage[T1]{fontenc}
\usepackage{dcolumn}
\usepackage{bm}
\usepackage{graphicx}
\usepackage{amsthm}
\usepackage{amsmath}
\usepackage{color}
\usepackage{xcolor}
\usepackage{verbatim}
\usepackage{esint}
\usepackage[english]{babel}
\usepackage{amsfonts}
\usepackage{slashed}
\usepackage{latexsym}
\usepackage{bm}
\usepackage[colorlinks = true,
            linkcolor = red,
            urlcolor  = blue,
            citecolor = magenta,
            anchorcolor = red]{hyperref}
\usepackage{esint}
\usepackage{soul}
\usepackage{cancel}
\usepackage[normalem]{ulem}
\usepackage{empheq}
\usepackage{dsfont}
\usepackage{amsmath}
\usepackage{epsfig} 
\usepackage{enumerate}
\definecolor{sapphire}{rgb}{0.03, 0.03, 0.41}

\DeclareMathAlphabet\mathbfcal{OMS}{cmsy}{b}{n}
\def\be{\begin{equation}}
\def\ee{\end{equation}}

\newcommand{\ket}[1]{| #1 \rangle}

\newcommand{\bpm}{\begin{pmatrix}}
\newcommand{\epm}{\end{pmatrix}}
\newcommand{\dg}{\dagger}
\newcommand{\la}{\langle}
\newcommand{\ra}{\rangle}
\newcommand{\sm}{\sigma}

\begin{document}

\title{Magnetism and superconductivity in doped triangular-lattice Mott insulators}

\author{Kun Woo Kim}
\affiliation{Department of Physics, Chung-Ang University, 06974 Seoul, Republic of Korea}
\author{T. Pereg-Barnea}
\affiliation{Department of Physics, McGill University, Montreal, Quebec H3A 2T8, Canada}

\date{\today}
\begin{abstract}
Inspired by recent advances in the fabrication of surface superlattices, and in particular the triangular lattice made of tin (Sn) atoms on silicon, we study an extended Hubbard mode on a triangular lattice. The observations of magnetism in these systems justify the inclusion of a strong on-site repulsion and the observation of superconductivity suggests including an effective, nearest-neighbor attractive interaction.  The attractive interaction mimics the effect of strong on-site repulsion near half filling, which can be seen in strong coupling vertex calculations such as the Eliashberg method.  With this extended Hubbard model on a triangular lattice with its geometrical frustration, we find a rich phase diagram of various magnetic orders and pairing functions, within the framework of self-consistent mean field theory.  We uncover the competition among magnetism and unconventional superconductivity, and their coexistence for triplet pairings. We follow the Fermi surface of the system as the system is doped away from half filling and find nesting vectors and a Lifshitz transition which provide an intuitive understanding of the phase transitions between the many orders we consider.
\end{abstract}
\maketitle

\section{Introduction}
\label{introduction}

Recent theoretical studies of superconductivity in triangular lattice systems have been motivated by a series of experimental investigation~\cite{ming2017realization,ming2018zero,wu2020superconductivity, ming2023evidence} where Sn adatoms  placed on the surface of Si(111) form a two-dimensional triangular lattice, showing some evidence for unconventional chiral d-wave superconductivity. The magnetic ordering of the same superlattice system has been intensively studied as well both theoretically and experimentally~\cite{PhysRevLett.98.126401,PhysRevLett.98.086401,li2011geometrical,li2013magnetic}.
These kind of systems combine the geometric frustration of the triangular lattice with strong electronic correlations coexist and can, at least in theory, bring about a variety of states of matter such as a spin liquid, collinear antiferromagnet and spiral magnetic order at half filling as well as several unconventional superconducting orders away from half filling. Here, we explore the competition between magnetism and superconductivity in a wide range of doping and interaction strengths. 

The studies of magnetic orders in a triangular lattice~\cite{davoudi2008competition, tocchio2014phase,PhysRevB.105.085102, willsher2023magnetic} and others investigate the possibility of unconventional superconductivity~\cite{cheng2010stable, wolf2018unconventional,jiang2020topological, wolf2022triplet, biderang2022topological} and point to the possibility of both singlet and triplet superconductors with and without topological numbers.  In this manuscript we focus on the competition between many order parameters, both magnetic and superconducting. We do so in self-consistent mean field theory where the on-site Hubbard $U$ interaction favors magnetic order while pairing on bonds is favored by an attraction term on nearest neighbor sites.  This attraction is an effective description of strong correlations; the result of a coulomb repulsion and fermi surface nesting~\cite{wolf2018unconventional}.  The tight binding model in the triangular model comes with the long range hoping sharpening the saddle of dispersion relation. We confirm not only the appearance of chiral d-wave superconductivity and collinear antiferromagnetic ordering as reported in experiments, but also find triplet pairing and the spiral magnetic ordering at other fillings and interaction strength. Crucially, we provide direct and intuitive understanding of magnetic phases from Fermi surface nesting and can relate the favored superconducting state to a synergy between the Fermi surface and the pairing function such that nodes are avoided and gaps are maximized.

The organization of the manuscript is the following. In section~\ref{sec:model} we introduce a model  with on-site repulsion and nearest neighbor attraction, and then the mean field Hamiltonian with magnetism and superconductivity is constructed. In section~\ref{sec:self_cons} we construct the grand potential and derive the self consistency relations for multiple order parameters. In sections~\ref{sec:results} and~\ref{sec:conclusions} we present our results and discuss them. 


\begin{figure*}
\centering
\includegraphics[width=0.9\textwidth]{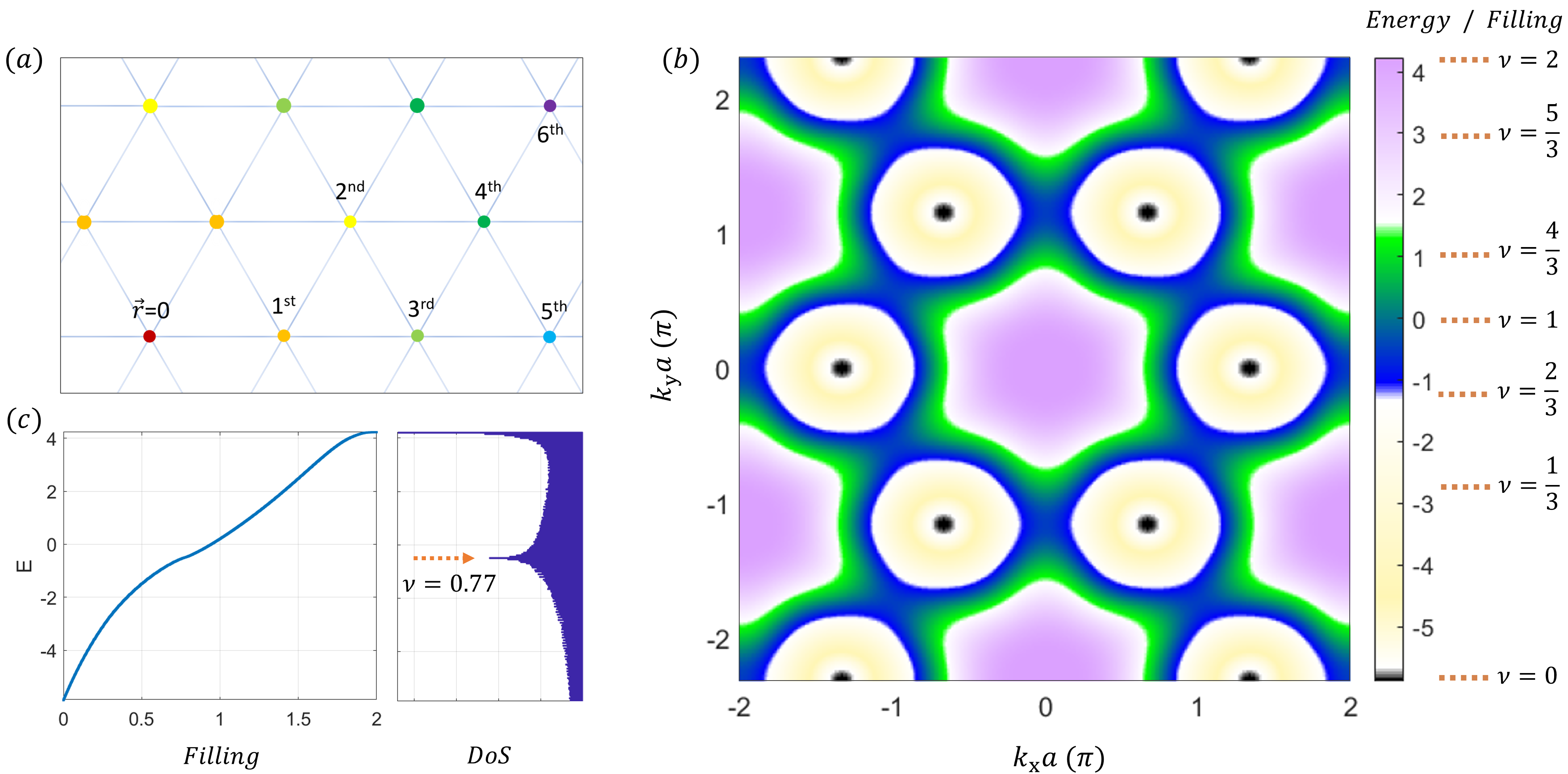}
\caption{\label{fig:2} (a) A triangular lattice with six nearest neighbors of the central atom (in red) marked with different colors.  (b) The dispersion relation $E(k_x,k_y)$ and the filling $\nu  = \la \hat n_{i\uparrow} + \hat n_{i\downarrow }\ra \equiv \la \hat n_i \ra$. Energy in unit of the nearest neighbor hopping amplitude, $|t_1|=52.8 (\text{meV})$. (c) The relation between filling and the energy, and the DoS and energy.}
\end{figure*}

\section{Extended Hubbard model on a triangular lattice}\label{sec:model}

\begin{figure}
\centering
\includegraphics[width=.45\textwidth]{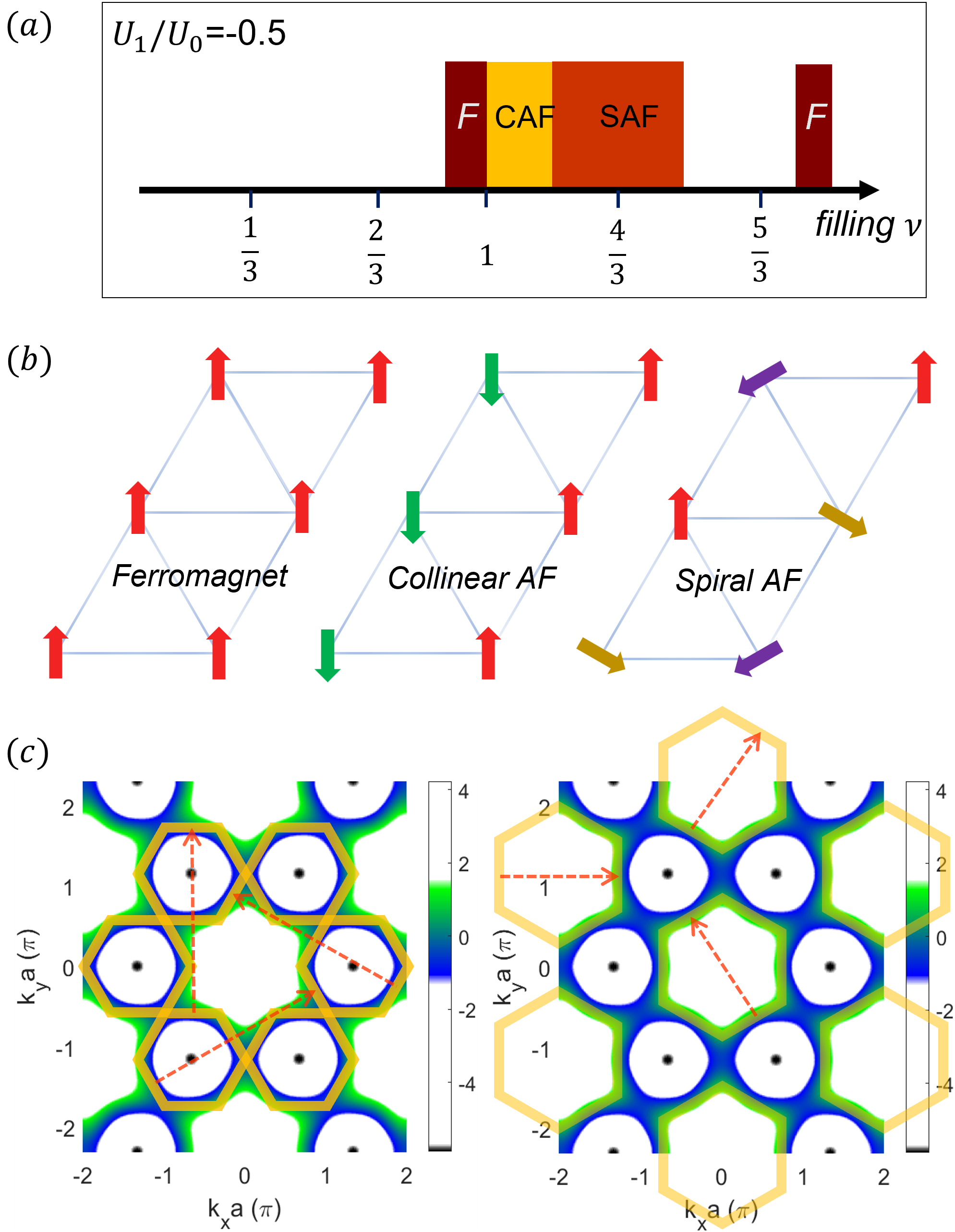}
\caption{\label{fig:3} (a) The phase diagram at $U_1/U_0=-0.5$. (b) Three types of magnetic orderings. (c) The collinear AF and spiral AF appear from the Fermi surface nesting depicted. The ferromagnetism appears from the Stoner's criterion. 
}
\end{figure}

The kinetic part of our Hamiltonian is composed of tight binding hopping parameters $t_{l}$ between the $l^\text{th}$ neighbors on the triangular lattice, proposed to match ARPES data in
Ref.~\cite{li2013magnetic}.  The parameters where chosen to fit the lowest energy band of the Sn/Si(111) surface:
\begin{align}
    \hat H_0 = \sum_{l=1}^6 t_l \sum_{\la ij\ra _l}  \hat c^\dagger_i \hat c_j, \label{tbm}
\end{align}
where $t_1=-52.7\text{meV}$, and the longer-range hopping amplitudes $t_2/t_1,..., t_6/t_1$ are -0.3881, 0.1444, -0.0228, 0 and -0.0318, respectively.   The indices $\langle i,j\rangle_l$ run over all pairs of $l$th nearest neighbor sites. The Fermi surface is depicted as a function of filling in Fig.~\ref{fig:2}(b) along with the density of states (DoS) as a function of energy.
We include the onsite and extended Hubbard interaction terms:
\begin{align}
    \hat H_{\text{int}} = \sum_i U_0 \hat n_{i\uparrow}\hat n_{i\downarrow} + \sum_{\la ij \ra,\sigma\sigma'} U_1 \hat n_{i\sigma}\hat n_{j\sigma'}, \label{hubbard}
\end{align}
such that $U_0>0$ is repulsive, and $U_1<0$ is attractive. 
The on-site repulsive interaction favors magnetism.  It is convenient to express the Hubbard $U_0$ term as (Appendix~\ref{sec:MF_hamiltonian}):
\begin{align}
    \hat H^{(\text{on})} = \sum_i U_0 \left[ \frac 1 4 \hat n_i \hat  n_i - \hat S_{im} \hat S_{im}\right], 
\end{align}
where $\hat n_i=\hat n_{i\uparrow} + \hat n_{i\downarrow}$ is the occupancy at site $i$, with $\langle \hat n_i \rangle \in [0,2]$. The spin operator at site $i$ in direction $\hat m$ is defined as $\hat S_{im} = \frac 1 2 \sum_{\sigma\sigma'}\hat c^\dagger _{i\sigma} (\vec \sigma \cdot \hat m)_{\sigma\sigma'} \hat c_{i\sigma'}$, where $\vec \sigma$ is the vector of Pauli matrices.
We can therefore define a local spin order parameter as $\vec m = \langle \hat S_{im} \rangle$, where the direction of $\vec m$ may give either ferromagnetic order or antiferromagnetic (AF) order with collinear or spiral spin directions as depicted in Fig.\ref{fig:3}(b).  The magnitude $m=|\vec m|$ represents the strength of the order.
The mean field Hamiltonian is then given by:
\begin{align}
\hat H^{(\text{on})}_{\text{MF}} = & -U_0 N_{\text{lat}} \left[ \frac{1}{4} \langle \hat n_i \rangle ^2 - |\vec m|^2 \right] \nonumber\\ & + U_0 \sum_i \bpm \hat c^\dagger _{i\uparrow} &  \hat c^\dagger_{i\downarrow} \epm  \left[ \frac 1 2 \langle \hat n_i \rangle \sigma_0 - \vec m \cdot \vec \sigma \right] \bpm \hat c_{i, \uparrow} \\\hat c_{i,\downarrow} \epm, \label{magH}
\end{align}
where the first term describes the onsite potential energy cost of having the magnetic order while the second term provides a possible energy benefit from a spin splitting. Note that the band energy shifts by $\frac 1 2\la \hat n_i \ra \sigma_0$ due to the mean repulsive  energy from the onsite interaction. 
To take into account the collinear and spiral AF magnetism we extended the unit cell to include 2 or 3 atoms with rotated spins in our calculations.

The attractive interaction $U_1<0$ can induce superconductivity or a charge density wave. We focus on superconductivity using BCS-like the self-consistent mean field:
\begin{align}
    \hat H^{(\text{nn})}_{\text{MF}} = &-U_1 \sum_{\langle ij\rangle,\sigma\sigma'}|\Delta_{ji,\sigma'\sigma}|^2 \nonumber \\ & + U_1 \sum_{\langle ij\rangle,\sigma,\sigma'} \left[ c_{i,\sigma}^\dagger c_{j,\sigma'}^\dagger \Delta_{ji,\sigma'\sigma} + \Delta ^* _{ij,\sigma\sigma'} c_{j\sigma'} c_{i\sigma}  \right]. 
\end{align}
The superconducting order parameter $\Delta_{ji,\sigma'\sigma}=\langle c_{j\sigma'}c_{i\sigma}\rangle$ may describe either singlet or triplet spin pairing with spatial symmetry of  $s$-, $p$-,$d$- or $f$-wave. We consider superconductivity and magnetism together to determine which combination of order parameters yields the lowest grand potential. It is worth noting that in order to consider superconductivity and magnetism simultaneously, the Hamiltonian in Eq.\eqref{magH} needs to be written in the Bogoliubov-de Gennes (BdG) form with particle-hole symmetry (see Appendix \ref{appendix:B}).  The mean field hamiltonian is 
\begin{align} 
   &\hat H_{\text{MF}} =  E_0(n_i , \vec m,\tilde\Delta^{(s/t)}_{ji,\sm'\sm}) && \nonumber \\ 
   &+\frac 1 2 \sum_k^{BZ/2} \Psi ^\dagger 
    \bpm H_{kk} & & & \tilde\Delta_{k,-k} \\ & H_{-k,-k}  & \tilde\Delta_{-k,k} & \\ & -\tilde \Delta^*_{k,-k} & -H_{kk}^* & \\ -\tilde\Delta^*_{-k,k}  & & &  -H_{-k-k}^* \epm \Psi , && \label{BdGH}
\end{align}
where  $\tilde\Delta_{k'k''} = U_1 \Delta_{k'k''}$ and $H_{k'k''}$ are 2 by 2 matrices in spin space. The BdG Hamiltonian is written in the basis $\Psi^\dagger = \left( c^\dagger_{k\uparrow}, c^\dagger_{k\downarrow}, c^\dagger_{-k\uparrow}, c^\dagger_{-k\downarrow}, c_{k\uparrow}, c_{k\downarrow}, c_{-k\uparrow}, c_{-k\downarrow} \right)$.  
The constant energy $E_0(n_i , \vec m,\tilde\Delta^{(s/t)}_{ji,\sm'\sm})$ contains the usual BCS ground state energy as well as terms resulting from the anti-commutation relation of the operators which compose the magnetic order parameters:
\begin{align}
    E_0 &= -U_0 N_{\text{lat}} \left[ \frac{1}{4} \langle \hat n_i \rangle ^2 - |\vec m|^2 \right] \nonumber \\ &+ \frac 1 2 U_0 N_{\text{lat}}\langle \hat n_i \rangle 
    -U_1 \sum_{\langle ij\rangle,\sigma\sigma'}|\Delta_{ji,\sigma'\sigma}|^2,
\end{align}

The Hamiltonian written in this structure visibly satisfies the particle hole symmetry which is represented by the operator $\mathcal P = K\tau_x$, where $K$ is complex conjugation and the Pauli matrix $\tau_x$ acts exchanges particles and holes,   
such that $\mathcal P H_{BdG} \mathcal P^{-1} = -H_{BdG}$ (see Appendix~\ref{appendix:B} for the discussion of the PHS in the basis $\Psi^\dg$). The block diagonal Hamiltonian in Eq.~\eqref{BdGH} is
\begin{align}
    H_{kk} = \left[\epsilon_{\vec k}+\frac{1}{2}U_0\la \hat n_i \ra \right]\sigma_0 - \vec m \cdot \vec \sigma, 
\end{align}
where $\epsilon_{\vec k}$ is the Fourier transform of Eq.~\eqref{tbm} (see \footnote{ 
$\epsilon_{\vec k}=-2t_1\left[ \cos(k_x) + 2\cos(\frac{\sqrt 3}{2}k_y) \cos(\frac{k_x}{2}) \right] \\ -2t_2 \left[  \cos(\sqrt 3 k_y) + 2\cos(\frac 3 2 k_x) \cos (\frac{\sqrt 3}{2} k_y ) \right] \\ - 2 t_3 \left[\cos (2k_x) + 2 \cos (k_x) \cos(\sqrt 3 k_y) \right] \\- 4 t_4 \left[ \cos (\frac 5 2 k_x) \cos (\frac{\sqrt 3}{2} k_y) + \cos(2k_x) \cos(\sqrt 3 k_y) \right. \\ \left. + \cos(\frac{k_x}{2}) \cos (\frac{3\sqrt 3}{2}k_y)  \right] - 2t_6 \left[ \cos(2\sqrt 3 k_y) +  2 \cos (3k_x) \cos (\sqrt 3 k_y) \right]$. where the fifth nearest neighbor hopping strength $t_5=0$.} for its explicit expression).
The summation over crystal momentum is on the half of the Brillouin zone because our basis contains both $\ket{k\sigma}$ and $\ket{\text{-}k\sigma}$ state. The BdG Hamiltonian (8 by 8) in the full basis is block diagonal with two 4 by 4 blocks with opposite sign of eigenvalues.  Note that one could choose to either work with this $8\times8$ block-diagonal Hamiltonian and sum over half of the Brillouin zone or work with only one of the blocks and sum over the entire Brillouin zone.
The pairing potential $\Delta_{k,-k}$ is given by:

\begin{align}
    (\Delta_{k,-k})_{\sm\sm'}=\sum_{\vec \delta_{ji}}  \Delta_{ji,\sm'\sm} e^{-i\vec k\cdot \vec\delta_{ji}}, \nonumber
\end{align}
where the sum is over the six nearest neighbor vectors, $\vec \delta_{ji}=\vec r_j-\vec r_i \in \{ (1,0), (\frac 1 2, \frac{\sqrt 3}{2}), (-\frac 1 2, \frac{\sqrt 3}{2}), (-1,0), (-\frac 1 2, -\frac{\sqrt 3}{2}), (\frac 1 2, -\frac{\sqrt 3}{2}) \}$, with angles $\theta_{ji} \in \{ 0, \frac{\pi}{3},\frac{2\pi}{3}, \pi, \frac{4\pi}{3}, \frac{5\pi}{3}  \} $ relative to the $x$-axis.
The symmetry of the pairing function in real space (odd or even with respect to exchanging the sites $i$ and $j$) determines whether the spins of the Cooper pair is in the singlet or triplet configuration such that $\Delta_{ji,\sigma\sigma'}$ is antisymmetric to the exchange of both position and spin.  
\begin{align}
\Delta_{ji,\sigma\sigma'} = \chi^{(\text{s/t})}_{\sigma\sigma'} \otimes \phi^{(\text{even/odd})}_{ji}. 
\end{align} 
and: 
\begin{align}
    \chi^{(s)} = \bpm 0  & +\Delta_s   \\ -\Delta_s & 0 \epm , \,\,\,    \chi^{(t)} = \bpm \Delta_{\uparrow\uparrow}  & \Delta_t   \\ \Delta_t & \Delta_{\downarrow\downarrow} \epm , 
\end{align}
We label the spatial pairing functions by their angular momentum; the phase winding number around a central atom such that for s-wave $\phi^{(s)}_{ji}=1$ and for angular momentum $l$ a chiral pairing function winds $l$ times and is given by $\phi^{(l)}_{jj'}\propto e^{il\theta_{jj'}}$. However, since we do not want to impose chirality a priori, we minimize the mean field energy with two pairing functions for each angular momentum:  
\begin{align}
&\phi^{(p_x)}_{ji}=\cos\theta_{ji}, \,\,\, &\phi^{(p_y)}_{ji}=\sin\theta_{ji}, \\
&\phi^{(d_{x^2-y^2})}_{ji}= \cos2\theta_{ji},\,\,\,&\phi^{(d_{xy})}_{ji}= \sin2\theta_{ji},\\
&\phi^{(f_{x(x^2-3y^2)})}_{ji}= \cos3\theta_{ji},\,\,\, &\phi^{(f_{y(3x^2-y^2)})}_{ji}= \sin3\theta_{ji},
\end{align}
such that the chiral $p$-, $d$-, or $f$-wave is obtained if both pairing functions are non-zero and there's a $\pi/2$ phase difference between them. Otherwise we end up with a non-chiral state.  In the case of $f$-wave, the $f_{y(3x^2-y^2)}$ is zero for nearest neighbor links in the triangular lattice and we therefore end up with a non-chiral, real order parameter of the form of $f_{x(x^2-3y^2)}$.  A chiral $f$-wave order would require longer range attraction.


\section{Self-consistency equations for order parameters}\label{sec:self_cons}

The grand potential is obtained from the grand canonical partition function $Z=\text{Tr}\, e^{-\beta(\hat H_{\text{MF}}-\mu \hat N)}$ in the diagonal basis.
\begin{align} \label{eq:H_diag}
\hat H_{\text{MF}}-\mu \hat N = E_0 + \frac{1}{2} \sum_{k,\alpha}^{BZ/2} \left[\zeta_{k,\alpha}\hat\gamma^\dg_{k\alpha}\hat\gamma_{k\alpha} -\zeta_{k,\alpha}\hat\gamma_{k\alpha}\hat\gamma^\dg_{k\alpha}\right], 
\end{align}
where $\alpha$ labels the eigen values and $\zeta_{k,\alpha}$ is an eigenvalue of one sub-block BdG Hamiltonian \eqref{BdGH}.  Due to particle hole symmetry, the eigenvalues in the two sub blocks of the Hamiltonian each have a pair of identical eigenvalues with opposite sign. Summing over all (single particle) eigen states we obtain the grand potential: 
\begin{align}
    \Omega = -\frac 1 \beta \ln Z  = E_0 - \frac 1 \beta \sum^{\frac 1 2 \text{BZ}}_{k,\alpha} \ln \left[  2 \cosh \frac{\beta \zeta_{k,\alpha}}{2}\right], 
\end{align}
where the temperature $\beta = k_BT=0.1|t_1|$ is taken and the sum above is only over positive $\zeta_{k,\alpha}$ values.  Note that this temperature of about 61K is required for convergence at our momentum resolution. The minimum of the grand potential is the self consistent solution for the order parameters and we therefore set the grand potential derivatives with respect to the order parameters to zero. We write the derivatives as (see Appendix~\ref{Sec:C}):
\begin{align}
    \frac{\partial \Omega}{\partial m_j} &= \frac{\partial E_0}{\partial m_j} + \frac{\partial}{\partial m_j} \left[ \Omega - E_0 \right] = 0, \\
    \frac{\partial \Omega}{\partial \Delta_\nu^{*}} &= \frac{\partial E_0}{\partial \Delta_\nu^{*}} + \frac{\partial}{\partial \Delta_\nu^{*}} \left[ \Omega - E_0 \right] = 0,
\end{align}
where $j\in \{x,y,z\}$ and $\nu\in \{s,t,\uparrow\uparrow,\downarrow\downarrow\}$.  
We solve the above conditions through iterations and update all order parameters at every step. Using the fact that the energy $E_0$ contains $\sum_{j=x,y,z}m_j^2$ and $\Delta_\nu\Delta_\nu^*$, we can update an order parameter through iteration. That is, 
\begin{align}
    m_{j}^{\text{(new)}} 
    &= m_{j}^{\text{(old)}} \left[ 1 - \eta \frac{\partial \Omega }{ \partial {m_j} } \left(\frac{\partial E_0}{\partial m_j}\right)^{-1} \right]_{\text{old}}, \\
    \Delta_{\nu}^{\text{(new)}} 
    &= \Delta_{\nu}^{\text{(old)}} \left[ 1 - \eta\frac{\partial \Omega }{ \partial \Delta^{*}_\nu } \left(\frac{\partial E_0}{\partial \Delta^{*}_\nu}\right)^{-1} \right]_{\text{old}}. 
\end{align}
where the right hand side is computed using a set of order parameters to be updated. $\eta \,(\simeq 0.2)$ controls the rate of approaching speed toward a convergence over iterations. When converged, $ m_{j}^{\text{(new)}}= m_{j}^{\text{(old)}}$ and $\left. \partial_{m_j} \Omega \right|_{\text{old}}=0$. 
The partial differentiation is numerically obtained by computing the difference of grand potentials, $\partial_{m_j} \Omega \simeq \lim_{\Delta m_j \rightarrow 0} \Delta \Omega/\Delta m_j$. The mean field Hamiltonian is also a function of filling $\nu $, see Eq.\eqref{magH}. Along with other order parameters, the filling is updated in each iteration  by computing
\begin{align}
    \nu = - \frac{1}{N_{\text{lat}}}\frac{\partial \Omega}{\partial \mu}=\la \hat n_{i\uparrow} + \hat n_{i\downarrow} \ra.
\end{align}

\section{Results: topological superconductivity and magnetism}\label{sec:results}

{\it Magnetic orders}--:  
The Stoner criterion gives a heuristic rule for when magnetic order might develop; if the density of states at the Fermi level exceeds a critical value set by the onsite repulsive interaction, magnetic order could develop to reduce the interaction energy.  This intuition is indeed in line with our findings - Fig.~\ref{fig:2} (c) shows that at filling near $\nu=0.77,2$ the DoS is peaked and this is where we find ferromagnetism. Another important factor in determining whether or not magnetism will appear and the kind of magnetism that will develop is the shape of the Fermi surface.  A magnetic order which reduces the periodicity is most effective if its ordering vector connects many states near the Fermi level, i.e., opens a gap through nesting. 


Around, $\nu \sim 1$, collinear antiferromagnetism appears. This order reduces the periodicity by folding the Brillouin zone in one of three directions; one such spin configuration is shown in Fig.~\ref{fig:3}(b). 
The periodicity of the collinear antiferromagnet (CAF)  in real space is $\sqrt{3}a$, which doubles the size of the unit cell. Thus, the size of the nesting vector $|\vec Q_{\text{CAF}}|=2\pi/\sqrt{3}a$ in momentum space is half of the shortest reciprocal lattice vector. The possible nesting Fermi surface lines are indicated in Fig.~\ref{fig:3}(c).  
However, since the CAF ordering appears only in one direction, the Fermi surface is not fully gapped by this order, and the system remains metallic. 
\begin{figure}
\centering
\includegraphics[width=.45\textwidth]{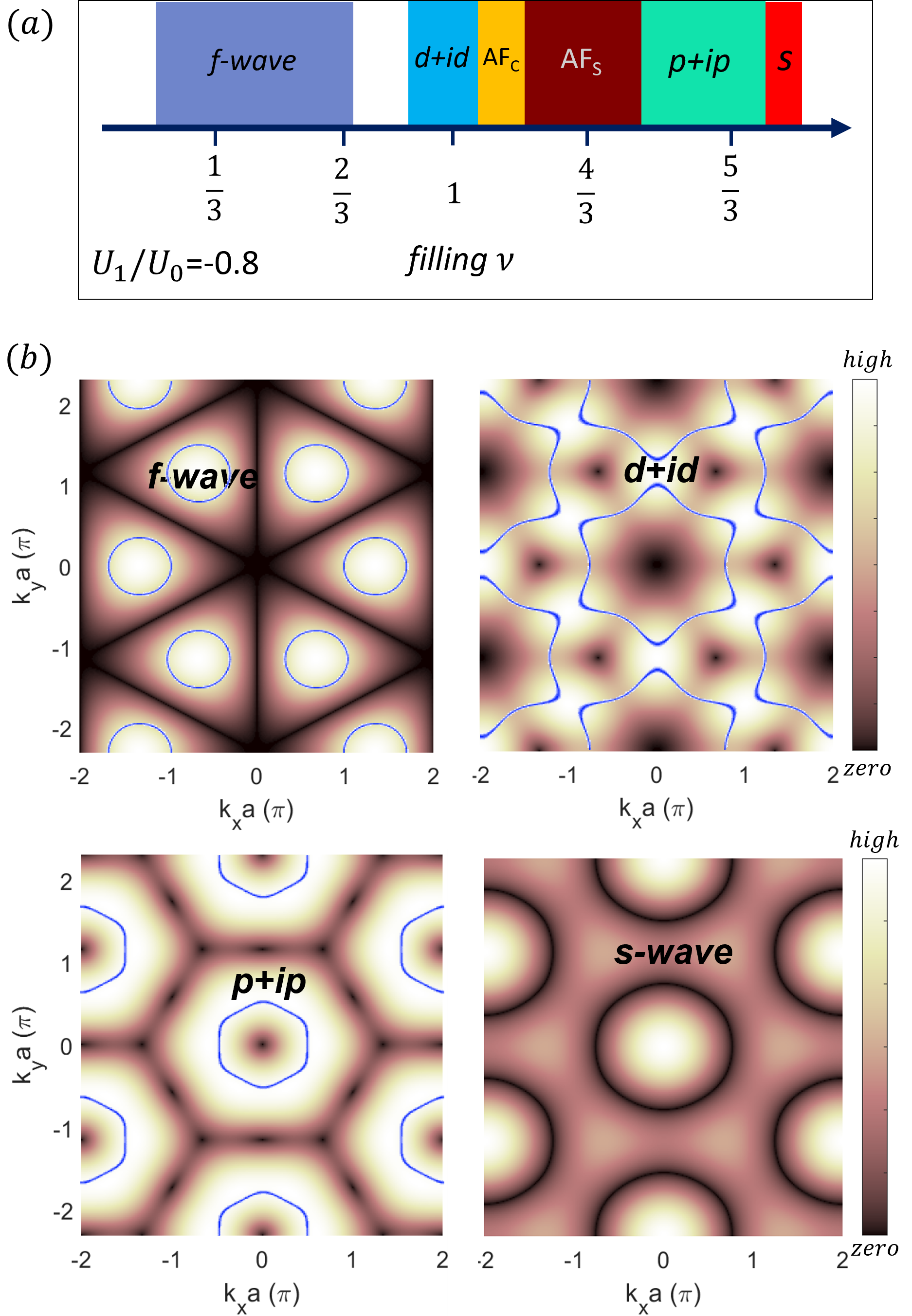}
\caption{\label{fig:4} (a) The phase diagram at $U_1/U_0=-0.8$. Chiral d-wave and p-wave superconductivity appear next to the anti-ferromagnetic orderings. At lower and higher filling, f-wave and s-wave superconductivity appear, respectively. (b) The magnitude of pairing  $|\Delta^{\text{(pairing)}}_{k,-k\sm\sm'}|$ is plotted for the orbital symmetries. The overlaid Fermi surface (blue lines) at filling $\nu=1/3,1,5/3$ show that nodal lines of the pairing are avoided maximizing the superconducting energy gap opening. }
\end{figure}

Right next to the CAF order, the spiral antiferromagnetic (SAF) is the most energetically favorable phase. The filling is a bit higher, and the corresponding Fermi surface is marked by green in Fig.~\ref{fig:3}(c). The size of the nesting vector is ${4\pi\over 3a}$ which is smaller than that of the CAF case, while the directions of the three vectors are rotated by $\pi/2$. The spiral AF gaps the Fermi surface completely, lowering the grand potential even more than the CAF. 

The development of the three magnetic orderings in a small doping range can be understood from the density of states (DoS) and the shape of the Fermi surface of the long range hopping tight binding model in the triangular lattice model for filling  $\nu \in [0.8, 1.5]$. The result of the self consistent mean field with multiple order parameters is shown in Fig.~\ref{fig:3} (a) for $U_0/|t_1|=6$ and $U_1/|t_1|=-3$. The bandwidth of the hopping model in Eq.~\eqref{tbm} ($\sim 10|t_1|$) as shown in Fig.\ref{fig:2} is the larger than employed interaction strength. We note that previous authors~\cite{li2013magnetic} found that the CAF is preferred over the SAF at half filling $\nu=1$ using a more sophisticated numerical method. When the attractive nearest neighbor attraction $U_1$ increases further, topological superconductivity appears next to the antiferromagnetic orders. 

{\it Superconductivity}--: The study of superconductivity in the one band triangular lattice models was ignited by the experimental observation of superconductivity in Sn/Si(111) \cite{ming2017realization,ming2018zero}. We therefore set out to find out what kind of superconductivity can emerge from our effective nearest neighbor attraction while on-site repulsion is present. This approach to interaction based unconventional superconductivity provides an intuitive understanding on how order parameters compete to balance decreasing the interaction energy while increasing the kinetic energy through orderings.  We therefore minimize our grand potential by considering both magnetism and superconductivity together.
\begin{figure*}
\centering
\includegraphics[width=1\textwidth]{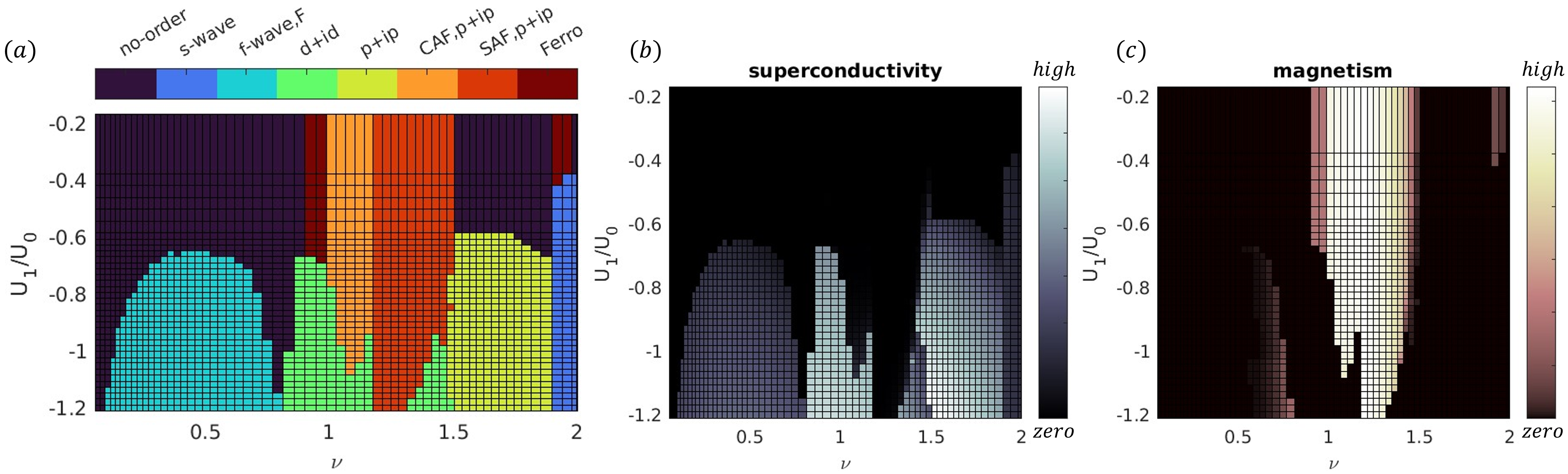}
\caption{\label{fig:5} (a) The phase diagram as a function of filling $\nu$ and the attractive interaction strength $U_1$ for $U_0=-6|t_1|$. (b,c) The corresponding superconducting and magnetic order parameters are plotted  in the same domain. They show the portion of order parameters for magnetism coexisting with triplet pairing superconductivity, p+ip and f-wave SC.}
\end{figure*}
Figure~\ref{fig:4}(a) shows the phase diagram as a function of filling with $U_1/U_0=-0.8$. The $f$-wave SC appears first and as can be seen in Fig.~\ref{fig:4}(b), it has three pairs of line nodes. For filling $\nu \in [0.2, 0.7]$, the Fermi surface is located where the nodal lines are avoided and therefore the $f$-wave pairing fully gaps the system, reducing the interaction energy which results from $U_1$. 

As the filling is increased beyond 0.7, the Fermi surface undergoes a Lifshitz transition from six pockets to a single snow-flake-shaped surface at the center of the Brillouin zone. Once the transition has occurred, the $f$-wave can not fully gap the Fermi surface and the preferred pairing changes. The Fermi surface that favors the CAF order for reducing the on-site interaction energy favors the chiral $d+id$ superconductor for reducing the $U_1$ interaction energy. The competition between the two phases is studied by minimizing the grand potential with the two orders present and this yields the result that $d+id$ superconductors appears before the collinear antiferromagnet. 

Around $\nu=3/4$ the spiral antiferromagnet is favored (despite the significant $U_1$) but at higher filling of $\nu>1.5$ another topological superconductor with $p+ip$ structure is developed. This order parameter has nodal points which are avoided by the small Fermi surface around the zone center as shown in Fig.~\ref{fig:4}(b)(bottom-left) such that the Fermi surface is fully gapped. 

Lastly, when every site is nearly fully filled, $\nu \sim 2$, the s-wave pairing potential which has a circular line node away from $\vec k=0$ is preferred. At the high filling, there is a transition from the ferromagnetic ordering to the $s$-wave superconductivity with increasing attractive extended interaction strength $|U_1|$, see Fig.~\ref{fig:5}(a). 

We note that the order of superconducting phases (chiral p-wave and f-wave) we find is inconsistent with that of Wolf {\it et al.}~in Refs.~\cite{wolf2018unconventional,wolf2022triplet} but it is consistent with that of Cheng {\it et al.}~in Ref.~\cite{cheng2010stable}.  While we rely on an effective interaction for our mean field calculation, we believe that the filling in which we find the various orders is very plausible since it is compatible with the intuition provided by the Fermi surface shapes at the various filling as discussed above.  

{\it From magnetic ordering to superconducting}--:
The phase diagram is drawn in Fig.~\ref{fig:5}(a) for $U_1/U_0 \in [-0.2, 1.2]$. At small $U_1$ only magnetic orders appear. For $U_1/U_1 <-0.4$ superconductivity begins to appear. While the singlet pairing does not coexist with the magnetic orderings, the triplet pairing may. We verify the coexistence and also plot the superconductor and the magnetic order parameters separately in Fig.~\ref{fig:5}(b) and (c), respectively. In particular $f$-wave superconductivity coexists with ferromagnetism near $\nu=0.75$, and the CAF and SAF orders give way to $p+ip$ pairing when $|U_1|$ is increased.



\section{Conclusions}\label{sec:conclusions}

In this work we explored the possibility of magnetic and superconducting orders in the extended Hubbard model on a triangular lattice.  Our model includes a repulsive on-site Hubbard interaction $U_0$ and an effective nearest neighbor attractive interaction $U_1$.  We treat the model with variational, self-consistent mean field theory which considers a large set of magnetic and superconducting orders together.  We map the phase diagram and find a ferromagnetic phase and two antiferromagnetic phases when the attractive interaction is weak.  for higher values of the attractive interaction we find superconducting states with $s$-, $d$-, $p$- and $f$-wave symmetry.  The $p$-wave and $d$-wave superconducting order parameters are found to be chiral/topological. 

Near filling $\nu=1-1.5$ the collinear anti-ferromagnetism and the spiral magnetism are consistent with previous studies~\cite{li2013magnetic}, are found to coexist with a $p+ip$ triplet topological superconductivity when the attractive interaction is significant. Our finding of $f$-wave and $p+ip$-wave superconductivity at low and high filling is also consistent with a previous study~\cite{cheng2010stable}, and we find  a $d+id$ topological superconductor to emerge when long range hopping is included in the kinetic energy. 

Our study has been inspired by recent advances in surface manipulation and in particular the creation of a triangular superlattice of tin atoms on a silicon surface~\cite{davoudi2008competition, tocchio2014phase, PhysRevB.105.085102, willsher2023magnetic}.  However, we expect these results to hold for other similar compounds.  In particular, our finding of chiral topological superconductivity could lead to the realization of Majorana zero modes at vortex cores in these compounds.  The proximity of these phases to other, non-topological phases suggests that the existence of Majorana modes could be controlled through gating and external fields. 

Lastly we would like to add a note about temperature.  Due to our limited momentum resolution we could not perform calculations at a very low temperature and used $\beta = k_B T = 0.1 |t_1| = 5.28 \text{(meV)}$ which is $\sim 61 \text{(K)}$.  This meant that in order to see gaps open we needed to work with large interaction values.  We therefore believe that at lower temperature one could see superconductivity with even weaker interactions.



{\it Acknowledgements -} K.W.K. acknowledges that this research was supported by the Chung-Ang University Research Grants in 2021. TPB acknowledges support from the Natural Sciences and Engineering Research Council of Canada (NSERC).

\appendix

\section{Mean-Field Decomposition of The Hubbard Interaction}\label{sec:MF_hamiltonian}

The onsite Hubbard interaction can be written as the density and spin operators using $\hat n_{i\uparrow}=\sum_{\alpha\beta}\frac 1 2 (1+\sigma_z)_{\alpha \beta} \hat c^\dg_{i\alpha}\hat c_{i\beta}$. More in general, the interaction can be written on the basis of arbitrary spin direction $\sigma_l = \hat l \cdot \vec \sigma$. Dropping the site index,  
\begin{align}
\hat n_{\uparrow}\hat n_{\downarrow} 
&=  \sum_{\alpha\beta\gamma\delta} \frac 1 4 (1+\sigma_l)_{\alpha\beta} (1-\sigma_l) _{\gamma\delta} \hat c^\dg_\alpha \hat c_\beta \hat c^\dg_\gamma \hat c_\delta,  \nonumber \\
&=  \sum_{\alpha\beta\gamma\delta} \frac 1 4 \left( \delta_{\alpha\beta}\delta _{\gamma\delta} - (\sigma_l)_{\alpha\beta} (\sigma_l)_{\gamma\delta}\right) \hat c^\dg_\alpha \hat c_\beta \hat c^\dg_\gamma \hat c_\delta, \nonumber 
\end{align}
where from the first line to the second the following relation is used: $\sum_{\alpha\beta\gamma\delta} (\sigma_l)_{\gamma\delta}\delta_{\alpha \beta} \hat c^\dg_\alpha \hat c_\beta \hat c^\dg_\gamma \hat c_\delta = \sum_{\alpha\beta\gamma\delta} (\sigma_l)_{\alpha\beta}\delta_{\gamma\delta}  \hat c^\dg_\gamma \hat c_\delta \hat c^\dg_\alpha \hat c_\beta $.
As a result, 
\begin{align}
\hat H^{\text{(on)}} &= U_0 \sum_i\left[ \frac 1 4 \hat n_i \hat n_i - \hat S_{il} \hat S_{il}\right],
\end{align}
where $\hat n_{i\sigma} = \hat c^\dg_{i\sm}\hat c _{i\sm}$, $\hat S_{il} = \frac{1}{2} \sum_{\alpha,\beta}\hat c^\dg_{i\alpha} (\hat l \cdot \vec\sigma)_{\alpha\beta} \hat c_{i\beta}$. 
Neglecting the fluctuating part of the interaction, the mean field approximation is
\begin{align}
    \hat n_i^2 &\simeq 2\hat n_i \la \hat n_i \ra - \la \hat n_i \ra ^2,\\
    \hat S_{il}^2&\simeq 2 \hat S_{il} \la \hat S_{il} \ra - \la \hat S_{il} \ra^2.
\end{align}
The mean field Hamiltonian \eqref{magH} is obtained. 

\section{Particle-Hole Symmetry and Grand Potential}\label{appendix:B}
The BdG Hamiltonian \eqref{BdGH} can be written as the sum of two sub blocks: 
\begin{align} \label{eq:bdg}
\hat H -\mu \hat N &= \frac 1 2 \sum_k^{\text{BZ/2}} \bpm \hat C^\dg_{k} & \hat C_{-k} \epm \bpm \tilde H_{k,k} & \tilde \Delta_{k,-k} \\ -\tilde \Delta_{-k,k}^* & -\tilde H_{-k,-k}^* \epm \bpm \hat C_{k} \\ \hat C_{-k}^\dg \epm \nonumber  \\
&+ \frac 1 2 \sum_k^{\text{BZ/2}} \bpm \hat C_{k} & \hat C^\dg_{-k} \epm \bpm -\tilde H_{k,k} & -\tilde \Delta_{k,-k} \\ \tilde \Delta^*_{-k,k} & \tilde H_{-k,-k}^* \epm^* \bpm \hat C_{k} \\ \hat C_{-k}^\dg \epm , 
\end{align}
where  $\hat C^\dg_k = \bpm c^\dg_{k\uparrow} & c^\dg _{k_\downarrow} \epm$ and $\tilde H_{k,k} = H_{k,k}-\mu$. 
The hermiticity of the sub block Hamiltonians are guaranteed because $\Delta^T = -\Delta$. By the diagonalization, 
\begin{align}
\bpm \tilde H_{k,k} & \tilde \Delta_{k,-k} \\ -\tilde \Delta_{-k,k}^* & -\tilde H_{-k,-k}^* \epm &= U \bpm \zeta_{1} & 0 \\ 0 & \zeta_{2} \epm U^\dg, \\
\bpm -\tilde H_{k,k} & -\tilde \Delta_{k,-k} \\ \tilde \Delta_{-k,k}^* & \tilde H_{-k,-k}^* \epm^* &= U^* \bpm -\zeta_{1} & 0 \\ 0 & -\zeta_{2} \epm U^T,
\end{align}
which verifies that the BdG Hamiltonian \eqref{BdGH} contains pairs of eigenvalues with the same magnitude and the opposite sign. Note that in general $\zeta_1 \neq -\zeta_2$ when $\tilde H_{k,k} \neq \tilde H_{-k,-k}$. Thus, within each sub block Hamiltonian, there is no particle hole symmetry as it is sometimes assumed in literature. The full spin and particle-hole basis must be employed for a system without inversion symmetry. In the eigen state basis, the Hamiltonian can be written as 
\begin{align}
\hat H -\mu \hat N&= \frac 1 2 \sum_k^{\text{BZ/2}} \bpm \hat \gamma_{k,1}^\dg & \hat \gamma_{k,2}^\dg \epm \bpm \zeta_{k,1} & 0 \\ 0 & \zeta_{k,2} \epm  \bpm \hat \gamma_{k,1} \\ \hat \gamma_{k,2} \epm \nonumber \\ 
&+ \frac 1 2 \sum_k^{\text{BZ/2}} \bpm \hat \gamma_{k,1} & \hat \gamma_{k,2} \epm \bpm -\zeta_{k,1} & 0 \\ 0 & -\zeta_{k,2} \epm  \bpm \hat \gamma_{k,1}^\dg \\ \hat \gamma_{k,2}^\dg \epm,
\end{align}
where $\bpm \hat \gamma_{k,1}^\dg & \hat \gamma_{k,2}^\dg \epm = \bpm \hat C_{k}^\dg & \hat C_{-k} \epm U $ and $U^* = U^{\dg T}$ are used. This is the mean field Hamiltonian Eq.~\eqref{eq:H_diag} used for the construction of the grand canonical partition function. 

Let us deduce the PHS relation that the block Hamiltonian matrix satisfies. The second term in Eq.~\eqref{eq:bdg} can be written as 
\begin{align} \label{eq:bdg2}
\hat H -\mu \hat N = \frac 1 2 \sum_k^{\text{BZ/2}} \bpm \hat C^\dg_{k} & \hat C_{-k} \epm \bpm \tilde H_{k,k} & \tilde \Delta_{k,-k} \\ -\tilde \Delta_{-k,k}^* & -\tilde H_{-k,-k}^* \epm \bpm  \hat C_{k}  \\ \hat C_{-k}^\dg  \epm \nonumber  \\
+ \frac 1 2 \sum_k^{\text{BZ/2}}  \bpm  \hat C^\dg_{-k} & \hat C_{k} \epm \tau_x \bpm -\tilde H_{k,k} & -\tilde \Delta_{k,-k} \\ \tilde \Delta^*_{-k,k} & \tilde H_{-k,-k}^* \epm^* \tau_x \bpm \hat C_{-k}^\dg \\ \hat C_{k}  \epm. 
\end{align}
The first term on the right hand side is the sum over momentum in one half of the Brillouin zone, and the second term is over the other half. And, the Hamiltonian in the second term is related to the first one by the following relation: 
\begin{align} \label{phs_bld}
    H_{\text{bld}}(-k) & = \tau_x \bpm -\tilde H_{k,k} & -\tilde \Delta_{k,-k} \\ \tilde \Delta^*_{-k,k} & \tilde H_{-k,-k}^* \epm^* \tau_x, \\
    & =- \tau_x \mathcal K H_{\text{bld}}(k) \tau_x \mathcal K, 
\end{align}
where subscript $_\text{bld}$ is to indicate the Hamiltonian matrices in Eq.~\eqref{eq:bdg2}. $H_\text{bdg}(k)$ is the Hamltonian matrix in the first term on the right hand side.

Next, let us deduce the PHS relation when the Hamiltonian matrix is constructed in the extended basis (also see pedagogical note in Ref.\cite{phs}) as in Eq.~\eqref{BdGH}. There is an unitary transformation between creation operators in real and momentum space:
\begin{align}
     \Psi_k= \bpm \vec C_k \\ \vec C^\dg_{k} \epm  = \bpm V &  \\ & V^* \epm \bpm \vec C_r \\ \vec C_r^\dg \epm. 
\end{align}
where the Fourier transformation $(V)_{ij}=\frac{1}{\sqrt{N_{\text{lat}}}}e^{-i \vec k_i \cdot \vec r_j}$ and 
\begin{align}
    \vec C_k &= (c_{k_1}, \cdots, c_{k_N})^T, \\
    \vec C^\dg_k &= (c^\dg_{k_1}, \cdots, c^\dg_{k_N})^T, \\
    \vec C_r &= (c_{r_1}, \cdots, c_{r_N})^T, \\
    \vec C^\dg_r &= (c^\dg_{r_1}, \cdots, c^\dg_{r_N})^T,
\end{align}
where spin and sublattice degree of freedom can be added in the operator vectors. Let $\Psi_r = \bpm \vec C_r \\ \vec C_r^\dg \epm$. Likewise, 
\begin{align}
     \Psi^\dg_k = \Psi_r^\dg \bpm V^\dg & \\ & V^{*\dg} \epm . 
\end{align}
Note that $\Psi_k^\dg$ and $\Psi_k$ take the same form with the one used in the mean field Hamiltonian, Eq.~\eqref{BdGH}.  The Hamiltonian  operator is
\begin{align}
    \hat H &=  \Psi^\dg_k H_k \Psi_k = \Psi_r^\dg \bpm V^\dg & \\ & V^{*\dg} \epm H_k \bpm V & \\  & V^* \epm \Psi_r ,\\
    &= -\Psi_r^\dg \tau_x \mathcal K \bpm V^\dg & \\ & V^{*\dg} \epm H_k \bpm V & \\  & V^* \epm \tau_x \mathcal K \Psi_r , \\
    &= -\Psi_r^\dg  \bpm V^\dg & \\ & V^{*\dg} \epm \tau_x \mathcal K H_k  \tau_x \mathcal K \bpm V & \\  & V^* \epm  \Psi_r , \\
    &= \Psi^\dg_k \tau_x (-H_k^*) \tau_x \Psi_k, 
\end{align}
where in the second line we used $H_r=-\tau_x \mathcal K H_r \tau_x \mathcal K$.  In the third line 
\begin{align}
    \tau_x \mathcal K \bpm V & \\  & V^* \epm \tau_x \mathcal K  = \bpm V & \\  & V^* \epm
\end{align}
is used. As shown in the last line,  the PHS relation in the momentum space is $H_k = - \tau_x \mathcal K H_k \mathcal K \tau_x$. This verifies that when Hamiltonian is written in the basis containing the creation and annihilation operators with the same set of indices as in $\Psi_{r}$ and $\Psi_{k}$, the particle-hole symmetry relation is equally applied in the form of    $\mathcal P H_r \mathcal P = -H_r$ and $\mathcal P H_k \mathcal P = -H_k$. Note the sign difference in momentum compared to Eq.~\eqref{phs_bld}, which applies to block diagonal Hamiltonian matrices in $H_k$.

\section{Self-Consistency Relations}
\label{Sec:C}
The mean field Hamiltonian includes order parameters whose values are determined self consistently. We start by some initial value for each order parameter and update the values at every iteration step in the following way.  With the 'old' set of order parameter values we calculate the mean field energies which are then used to calculate the grand potential.  The 'new' set of order parameter values are then calculated from the grand potential and the process is continued until conversions, when the values do not change between iterations.

Calculating the order parameter from the grand potential can be demonstrated for ferromagnetic order in the $z$ direction, $m_z$. Differentiating the grand potential by $m_z$ gives, 
\begin{align}
{d \Omega \over dm_z} = \frac{\partial (\Omega-E_0)}{\partial m_z} + {d E_0 \over dm_z}, 
\end{align}
where $\partial_{m_z} E_0 = U_0 N_{\text{lat}} \left( 2m_z \right)$, and 
\begin{align}    
\frac{\partial (\Omega-E_0)}{\partial m_z} &= -\beta^{-1} \frac{1}{ Z} \frac{\partial Z}{\partial m_z} \nonumber \\ &= -\frac{U_0}{Z} \text{Tr} \left[ \left( \sum_i \hat c^\dg_{i\uparrow}\hat c_{i\uparrow}- \hat c^\dg_{i\downarrow}\hat c_{i\downarrow}\right)e^{-\beta (\hat H- \mu \hat N )} \right] \nonumber \\
& \equiv -U_0 \la \sum_i^{N_{lat}}  \hat c^\dg_{i\uparrow}\hat c_{i\uparrow}- \hat c^\dg_{i\downarrow}\hat c_{i\downarrow}   \ra , 
\end{align}
Therefore, when the minimum of the grand potential is found, $\partial_{m_z} \Omega = 0$, 
\begin{align}
m_z = \frac{1}{2} \la  \hat c^\dg_{i\uparrow}\hat c_{i\uparrow}- \hat c^\dg_{i\downarrow}\hat c_{i\downarrow}   \ra, 
\end{align}
which is consistent with the way $\hat S_{iz}$ operator is defined in Eq.(3). For CAF and the SAF, we prepare the order parameter in an enlarged unit cell. The relative spin angle within the unit cell is arranged in such a way that it realizes the desired spin ordering and the strength of the magnetism is the order parameter which is found self-consistently. 

The self-consistent relation for the superconducting order parameter similarly follows. Let us begin from the real space expression Eq.(5). 
\begin{align}
\frac{\partial (\Omega-E_0)}{\partial \Delta_{ji,\sm'\sm}} &= \frac{U_1}{Z} \text{Tr} \left[\left( \hat c^\dg_{i\sm}\hat c^\dg_{j\sm'} \right) e^{-\beta (\hat H - \mu \hat N)} \right], \\
& =  U_1 \la \hat c^\dg_{i\sm}\hat c^\dg_{j\sm'} \ra, 
\end{align}
where we pick a specific site index $i,j$ for the pairing order parameter, hence there is no summation. Since $\partial_{\Delta_{ji,\sm'\sm}} E_0 = -U_1 \Delta_{ji,\sm'\sm}^*$ we arrive at, 
\begin{align}
\Delta_{ji,\sm'\sm}^* = \la \hat c^\dg_{i\sm}\hat c^\dg_{j\sm'} \ra. 
\end{align}
Note the positions of indices, $(\Delta^\dg)_{ij,\sm\sm'} = \Delta_{ji,\sm'\sm}^*$. For a general search of superconductivity with a certain pairing symmetry, we transform the Hamiltonian to momentum space using $\hat c_{i\sm}^\dg =\frac{1}{\sqrt{N_{\text{lat}}}} \sum_k e^{ikr_i}\hat c_{k\sm}^\dg$: 
\begin{align}
H_{\text{sc}} = U_1\sum^{\text{BZ}}_{k,\sm\sm'\delta} \Delta_{\delta,\sm'\sm} \hat c^\dg_{k\sm}c^\dg_{-k\sm'}e^{-ik\delta}+h.c.
\end{align}
where $\delta$ goes over the six nearest neighbor vectors. The summation in momentum is then divided by half. This is to prepare the BdG Hamiltonian where the basis $\Psi^{(\dg)}$ includes both $k$ and $-k$. 
\begin{align}
H_{\text{sc}} &= U_1\sum^{\text{BZ/2}}_{k,\sm\sm'} \left(\sum_{\delta}\Delta_{\delta,\sm'\sm}e^{-ik\delta}\right) \hat c^\dg_{k\sm}c^\dg_{-k\sm'} \nonumber \\
& + U_1\sum^{\text{BZ/2}}_{k,\sm\sm'} \left(\sum_{\delta}\Delta_{\delta,\sm'\sm}e^{ik\delta}\right) \hat c^\dg_{-k\sm}c^\dg_{k\sm'} +h.c.
\end{align}
where the bracket in the first and the second term are $\Delta_{k-k,\sm\sm'}$ and $\Delta_{-kk,\sm\sm'}$ in the BdG Hamiltonian Eq.\eqref{BdGH}. The order parameter $\Delta_{\delta,\sm'\sm}= \chi_{\sm'\sm} \phi_{\delta} $. Where the spin configuration is in $\chi_{\sm,\sm'}$ and the desired order parameter structure is $\phi_{\delta}$ which encodes a uniform order parameter whose magnitude and phase may depend on the bond direction.  The magnitude of the order parameter is included in $\chi_{\sm'\sm}$ and we therefore find it iteratively
by performing the differentiation of the grand potential, 
\begin{align}
&\frac{\partial (\Omega- E_0)}{\partial \chi_{\sm'\sm}} \nonumber \\ &=
\frac{U_1}{Z} \text{Tr} \left[\left(\sum_{k\delta}^{\text{BZ}}\phi_{\delta}e^{-ik\delta} \hat c^\dg_{k\sm}\hat c^\dg_{-k\sm'} \right) e^{-\beta (\hat H - \mu \hat N)} \right], \nonumber \\ 
&=U_1  \sum_{k\delta}^{\text{BZ}}\phi_{\delta}e^{-ik\delta} \la\hat c^\dg_{k\sm}\hat c^\dg_{-k\sm'}  \ra , \\
&=U_1  \sum_{k\delta}^{\text{BZ}}\phi_{\delta}e^{-ik\delta} \frac{1}{N_{\text{lat}}}\la \sum_{i} e^{-ikr_i}\hat c^\dg_{i\sm} \sum_{j} e^{ikr_j} \hat c^\dg_{j\sm'}  \ra ,
\end{align}
where in the last line we bring it back to real space expression using $\hat c_{k\sm}^\dg = \frac{1}{\sqrt{N_{\text{lat}}}}\sum_{i} e^{-ikr_i}\hat c_{i\sm}^\dg$. the summation over momentum yields the delta function $\delta (r_j - r_i - \delta)$. As a result, 
\begin{align}
\frac{\partial (\Omega- E_0)}{\partial \chi_{\sm'\sm}} \nonumber 
&= U_1 \sum_{r_i\delta} \phi_{\delta} \la \hat c^\dg_{i\sm}  \hat c^\dg_{i+\delta_\sm'}  \ra , \\
&= U_1 N_{\text{lat}} \sum_{\delta} \phi_{\delta} \la  \hat \Delta^\dg_{\delta,\sm\sm'}  \ra , \nonumber \\
&= U_1 N_{\text{lat}}\la  \hat \chi^\dg_{\sm\sm'}  \ra\sum_{\delta} |\phi_{\delta}|^2 , 
\end{align}
where the summation over momentum is restored to the whole BZ by combining $k$ and $-k$ terms. Because the constant energy term provides 
\begin{align}
\partial_{\chi_{\sm'\sm}} E_0 &= -U_1  \sum_{ij} \partial_{\chi_{\sm'\sm}} |\Delta_{\delta,\sm'\sm}|^2, \nonumber \\
&=-U_1\chi_{\sm'\sm}^* N_{\text{lat}}\sum_{\delta}
|\phi_\delta|^2. 
\end{align}
where $\Delta_{\delta,\sm'\sm}=\chi_{\sm'\sm}\phi_{\delta}$ is used in the first line. Therefore, at the minimum of the grand potential, when $\partial_{\chi_{\sm'\sm}} \Omega=0$ we have,
\begin{align}
\chi^*_{\sm'\sm} = \la  \hat \chi^\dg_{\sm\sm'}  \ra,
\end{align}
where $\la  \hat \chi^\dg_{\sm\sm'}  \ra=\la  \hat \chi^*_{\sm'\sm}  \ra$.  Therefore, 
\begin{align}
\Delta^\dg_{\delta,\sm\sm'} = \la \hat \Delta^\dg_{\delta,\sm\sm'} \ra,
\end{align}
This verifies the superconductivity self consistency relation.






\bibliography{biblio}
\end{document}